\pdfoutput=1 
\documentclass{JINST-Sample-files/JINST}

\usepackage{url}
\usepackage{graphicx}
\usepackage{caption}
\usepackage{subcaption}

\title{Achievements of the ATLAS Upgrade Planar Pixel Sensors R\&D Project}

\author{C. Nellist$^a$\thanks{Corresponding author.}~,
on behalf of the PPS Collaboration\\
\llap{$^a$}Laboratoire de l'Accélérateur Linéaire, CNRS\\
Bat. 200, 9140, Orsay, France.\\
E-mail: \email{clara.nellist@cern.ch}}

\abstract{In the framework of the HL-LHC upgrade, the ATLAS experiment plans to introduce an all-silicon inner tracker to cope with the elevated occupancy.

To investigate the suitability of pixel sensors using the proven planar technology for the upgraded tracker, the ATLAS Planar Pixel Sensor R\&D Project (PPS) was established comprising 19 institutes and more than 90 scientists. The paper provides an overview of the research and development project and highlights accomplishments, among them: beam test results with planar sensors up to innermost layer fluences (~>~$10^{16}~$n$_{eq}$~cm$^{-2}$); measurements obtained with irradiated thin edgeless n-in-p pixel assemblies; recent studies of the SCP technique to obtain almost active edges by post-processing already existing sensors based on scribing, cleaving and edge passivation; an update on prototyping efforts for large areas: sensor design improvements and concepts for low-cost hybridisation; comparison between Secondary Ion Mass Spectrometry results and TCAD simulations. Together, these results allow an assessment of the state-of-the-art with respect to radiation-hard position-sensitive tracking detectors suited for the instrumentation of large areas.}


\begin{document}

\section{Introduction}\label{sec:intro}

The Large Hadron Collider (LHC)~\cite{ref:LHC} at CERN collides protons together at the centre of experiments with the aim of studying the fundamental properties of particles and the forces that bind them. ATLAS~\cite{ref:ATLAS} is one of two general purpose detectors on the LHC which announced the discovery of a new particle in 2012~\cite{ref:Higgs}, consistent with the predictions of a Standard Model (SM) Higgs boson. In order to expand the physics programme at CERN, three shutdowns for upgrades have been scheduled, with each period of data-taking known as a `run' and increasing sequential numbers after each long-shutdown. This paper focuses on the preparations for the upgrade of the pixel detector for the ATLAS experiment.

The ATLAS detector is constructed from a number of sub-detector layers which have been designed to measure the properties of particles produced from collisions at the centre of the detector. The structure consists of barrels lying parallel to the beam-pipe which fit inside each other, and end-caps oriented perpendicularly to the beam-pipe, ensuring continuous data coverage by completing the high angle regions.

\subsection{Pixel Detector of Run I}\label{sec:RunI}

\begin{center}
\begin{figure}[tbp]
\begin{subfigure}[c]{.56\textwidth}
\includegraphics[width=\textwidth]{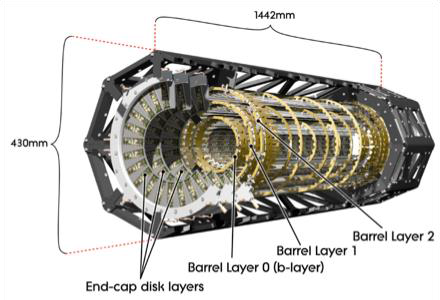}
\vspace{1.6mm}
\caption{}
\label{fig:Run1:ATLASPixelDetector}
\end{subfigure}
\begin{subfigure}[c]{.43\textwidth}
\includegraphics[width=\textwidth]{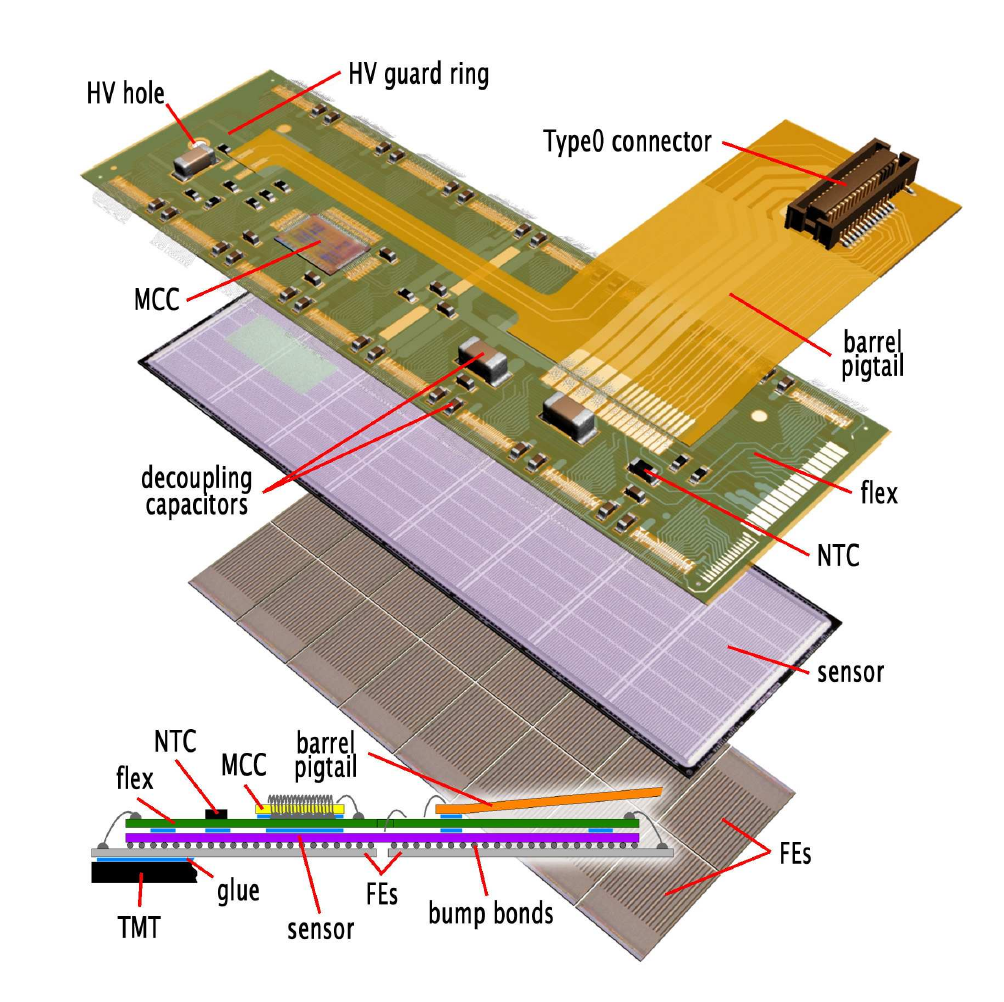}\\
\caption{}
\label{fig:Run1:FEI3}
\end{subfigure}
\caption{(a) The ATLAS pixel sub-detector of Run I. The three barrel layers and end-cap disk layers are clearly labelled and the scale is given~\cite{ref:ATLASImage}. (b) An ATLAS FE-I3 Barrel Pixel Module for pixel devices installed for the Run I data taking period. The hybrid structure is made up of sixteen front-end chips, the silicon sensor tile, the flex and the barrel pigtail~\cite{ref:FEI3Module}.}
\end{figure}
\end{center}

The aim of the ATLAS pixel sub-detector, shown in figure~\ref{fig:Run1:ATLASPixelDetector}, is to provide high granularity track points in order to allow the reconstruction of particle positions as accurately as possible. The original ATLAS pixel detector design consisted of three barrel layers at 50.5~mm, 88.5~mm and 122.5~mm from the proton-proton interaction point respectively and three disks at either end. The sensor design for these modules is a planar n$^+$-in-n silicon pixel sensor with a bulk thickness of 256~$\mu$m. Each sensor is bump-bonded to a set of ATLAS FE-I3 front-end read-out chips \cite{ref:FEI3}, this is known as a hybrid module. A complete ATLAS barrel pixel module of sixteen front-end chips, the silicon sensor tile, the flex and the barrel pigtail, is illustrated in figure~\ref{fig:Run1:FEI3}~\cite{ref:FEI3Module}. The FE-I3 read-out chip has a standard pixel size of 50~$\mu$m~by~400~$\mu$m and one chip consists of an array of 160~rows by 18~columns. Located at the edge of the chip are extended pixels of 50~$\mu$m by 600~$\mu$m to ensure complete coverage between neighbouring front-end chips.

\subsection{Pixel Detector of Run II}\label{sec:RunII}
After the first long shut-down between 2013 and 2014, the collision energy of the LHC will increase to the design energy of ${14}$~TeV. There will also be a higher luminosity of up to 1--1.7~x~10$^{34}$, resulting in an increase of pile-up within the pixel detector. Furthermore, the proximity to the proton collisions means that the the pixel layers will receive the highest radiation dose of all ATLAS sub-detectors. Consequently, to ensure that full tracking capability was maintained, a fourth pixel layer was added between the current pixel detector and a new, smaller beam pipe. This additional layer is known as the Insertable B-Layer, or IBL~\cite{ref:IBL}, and has a radius of $\sim$ 30~mm. A new ATLAS front-end read-out chip, called FE-I4, was designed and produced with a smaller pixel size of 50~$\mu$m by 250~$\mu$m~\cite{ref:FE-I4}.
 
 A mixed sensor technology was chosen for the 14 IBL staves. Planar pixel double-chip modules with slim edges were selected to populate the centre of each stave, while the 3D silicon~\cite{ref:3D} single-chip modules were chosen for the high-eta regions.
 
The IBL was successfully installed in May 2014 and will begin taking data when the LHC resumes collisions in early 2015, resulting in improved tracking precision~\cite{ref:IBL2}.

\subsection{Phase II Upgrade}\label{sec:PhaseII}

The phase II upgrade of the LHC, known as the High-Luminoscity LHC (HL-LHC) is foreseen to occur during the years 2022--2024 with a new period of data taking beginning in 2025. The expected luminosity after the upgrade is 5~x~10$^{34}$~cm$^{-2}$~s$^{-1}$ with 3000 fb$^{-1}$ of expected data collected at ATLAS; this is required for the precision measurements of the properties of the Higgs boson and physics beyond the SM. At the end of the data-taking period a fluence of $\sim$~2~x~10$^{16}$~n$_{eq}$~cm$^{-2}$ is expected in the inner-most layer of the ATLAS detector (at $\sim$~4~mm from the centre), where 1~n$_{eq}$ is 1~MeV neutron equivalent fluence. As a consequence, improved pixel devices are required to cope with this extreme environment. 

\begin{figure}[tbp] 
\centering
\includegraphics[width=0.7\textwidth]{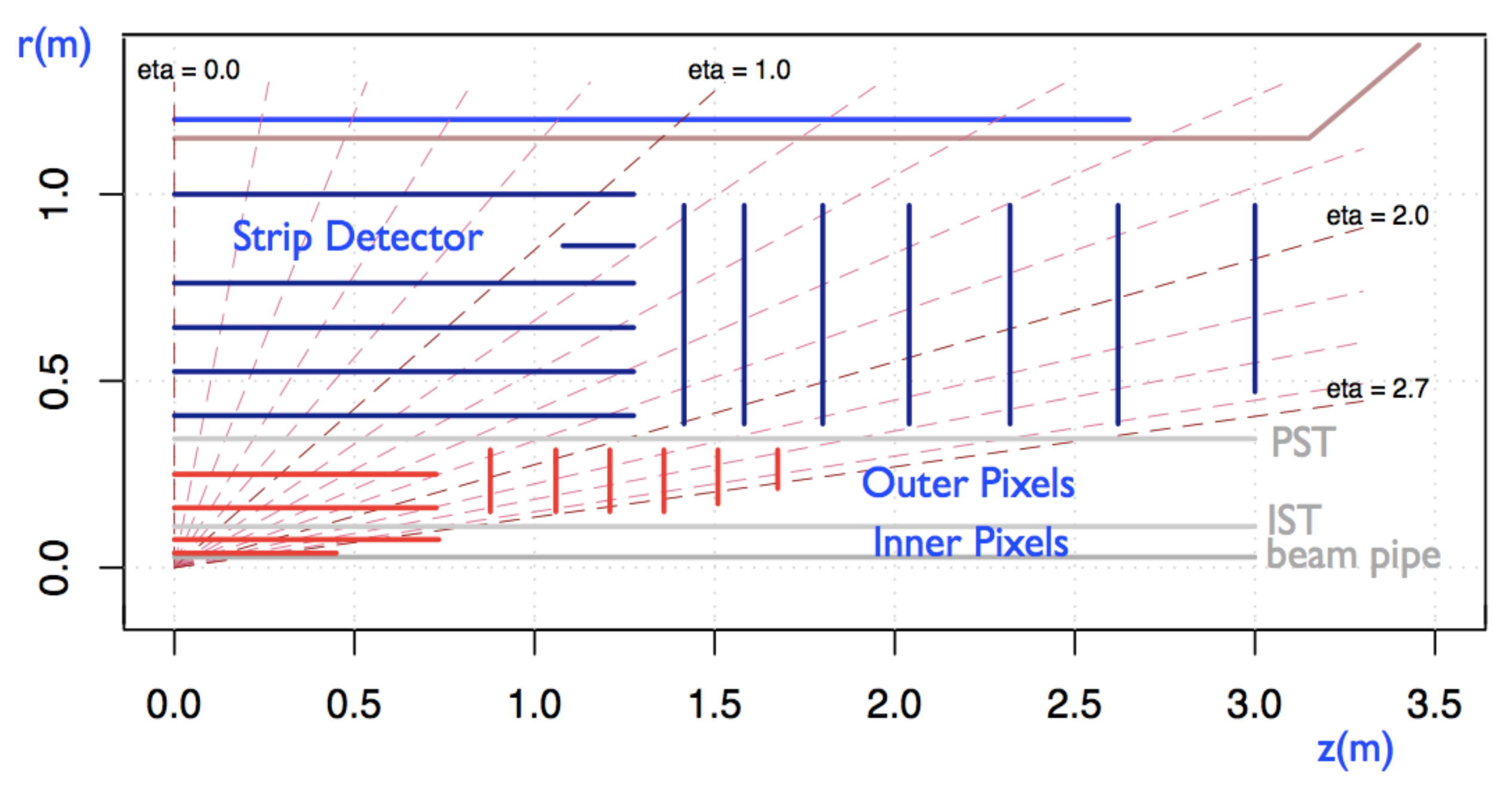}
\caption{Diagram of the foreseen all-silicon Inner Tracker (ITk) upgrade of the ATLAS Inner Detector, due to be installed in the phase II upgrade.}
\label{fig:PhaseII:ITk}
\end{figure}

For ATLAS, this will also mean a new, entirely silicon tracker with a greater area of pixel detectors from 1.7 m$^{2}$ to approximately 8~m$^{2}$ called the Inner Tracker (ITk)~\cite{ref:PhaseIIUpgrade}. This is due to a higher coverage in the eta region and also a larger radius of the outer pixel layer with respect to the present layout. Figure~\ref{fig:PhaseII:ITk} shows a baseline layout of this new all-silicon inner-detector, including the inner-pixel layers, the outer-pixel layers and the strip detector region respectively as the distance from the interaction point, at (0.0, 0.0) in the diagram, increases.

For the inner layers of the pixel detector, the main areas of focus in the R\&D activities are radiation hard devices,
slimmer or active edges, and better granularity (smaller pixels), while for the outer layers the achievements are low-cost and high-yield production techniques for multi-chip modules.

\paragraph{Planar Pixel Sensors}

The following section describes recent achievements by the ATLAS  Planar Pixel Sensor (PPS) collaboration, part of the ITk pixel group, in working towards this goal.

\section{PPS Results}\label{sec:PPSresults}
\subsection{Test Beam Results for Inner Layers}\label{sec:BTResults}

\paragraph{Test Beam Setup}

It is vital to study prototype devices in an experiment that mimics the environment they will be exposed to in ATLAS.  Several beam tests of planar pixel devices have taken place since 2012 at various facilities. Results presented below have been collected from test beams at the CERN SPS facility using 120~GeV muons; at SLAC, USA using a 12.5~GeV e$^{-}$ beam; and at DESY using a 4~GeV e$^{+}$ or e$^{-}$ beam.

Particle tracks through the prototype devices are reconstructed using the EUDET~\cite{ref:EUDET} telescope hardware and EUTelescope~\cite{ref:EUTelescope} offline reconstruction software. The EUDET telescope consists of six Mimosa26~\cite{ref:Mim} planes, three-upstream and three-downstream of the central testing area. The EUTelescope reconstruction software uses MILLIPEDE~\cite{ref:Milli} to align the devices during the reconstruction process. The devices under test are positioned in the central area and readout using either the USBPix system~\cite{ref:USBPix} or the RCE system~\cite{ref:RCE}.

\paragraph{Results}

\begin{figure}[tbp]
\centering
\begin{subfigure}[b]{0.35\textwidth}
\includegraphics[width=\textwidth]{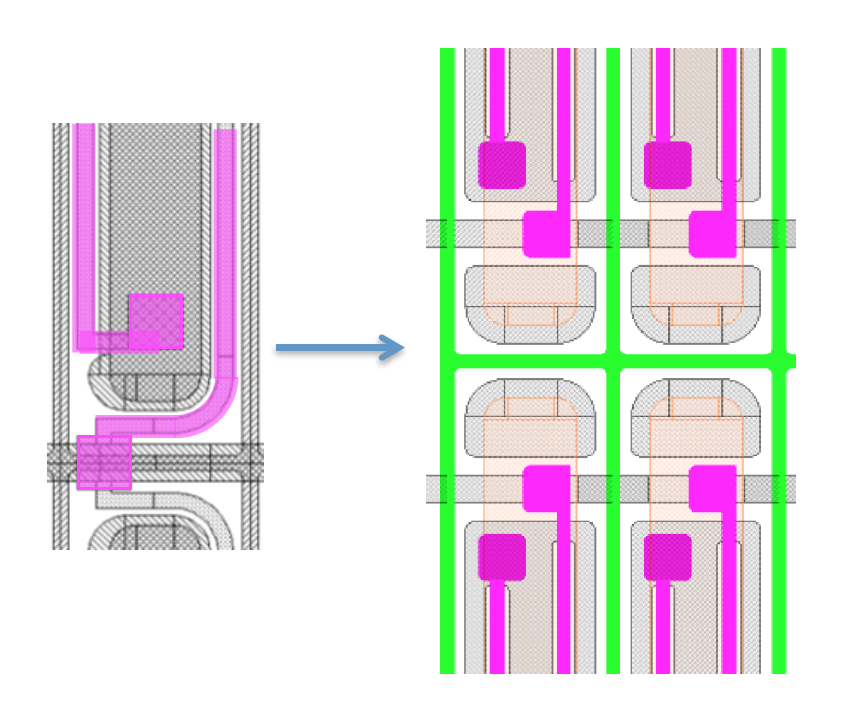}
\caption{}
\label{fig:BR}
\end{subfigure}
\begin{subfigure}[b]{0.55\textwidth}
\includegraphics[width=\textwidth]{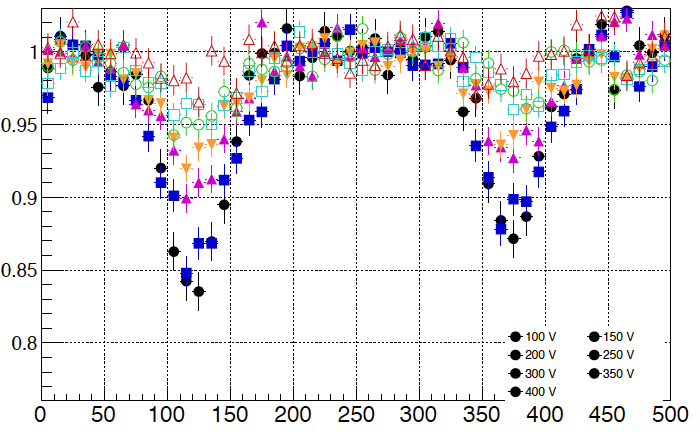}
\caption{}
\label{fig:BT:KEK}
\end{subfigure}
\caption{(a) Illustration of the previous (left) and new (right) pixel cell design. The PolySi has been moved inside the pixel to mask the ground potential. (b) Measured efficiency of a FE-I4 module produced at HPK, known as KEK46, projected along the length of a single pixel cell.}

\end{figure}

\begin{center}
\begin{figure}[tbp] 
\begin{subfigure}[b]{0.51\textwidth}
\includegraphics[width=\textwidth]{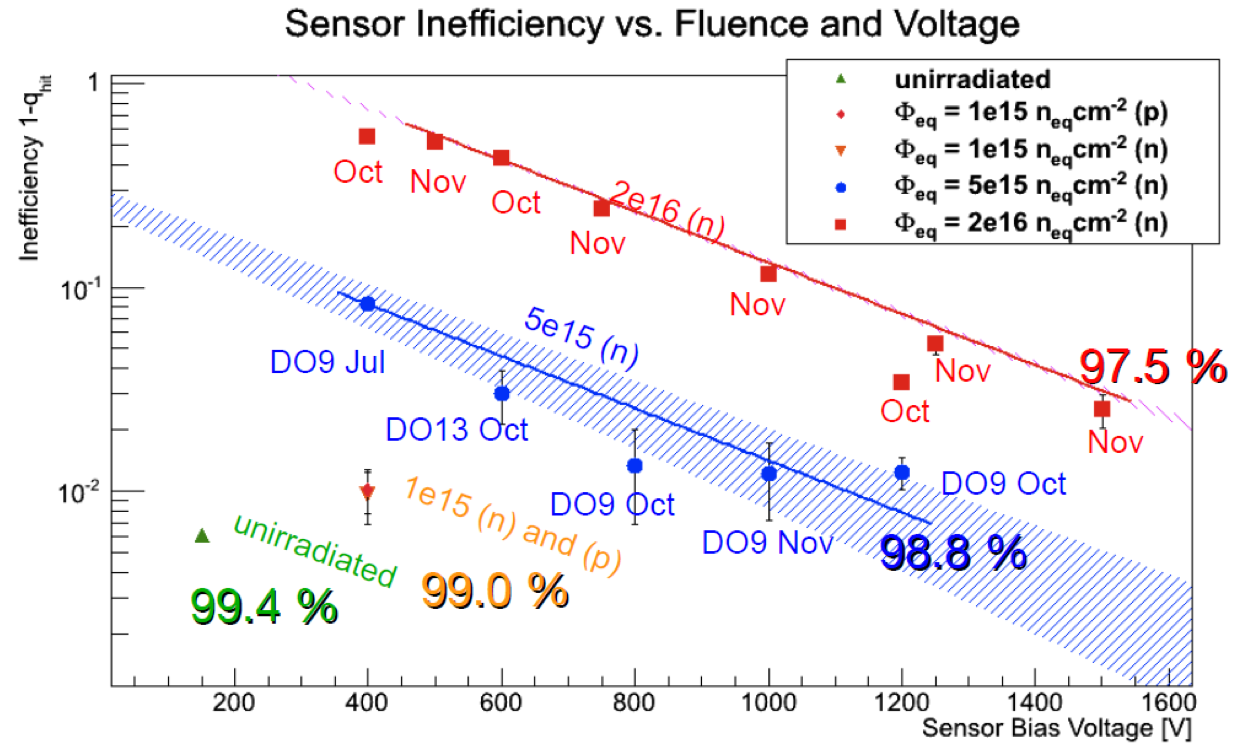}
\caption{}
\label{fig:BT:ineff}
\end{subfigure}
 \begin{subfigure}[b]{0.49\textwidth}
\includegraphics[width=\textwidth]{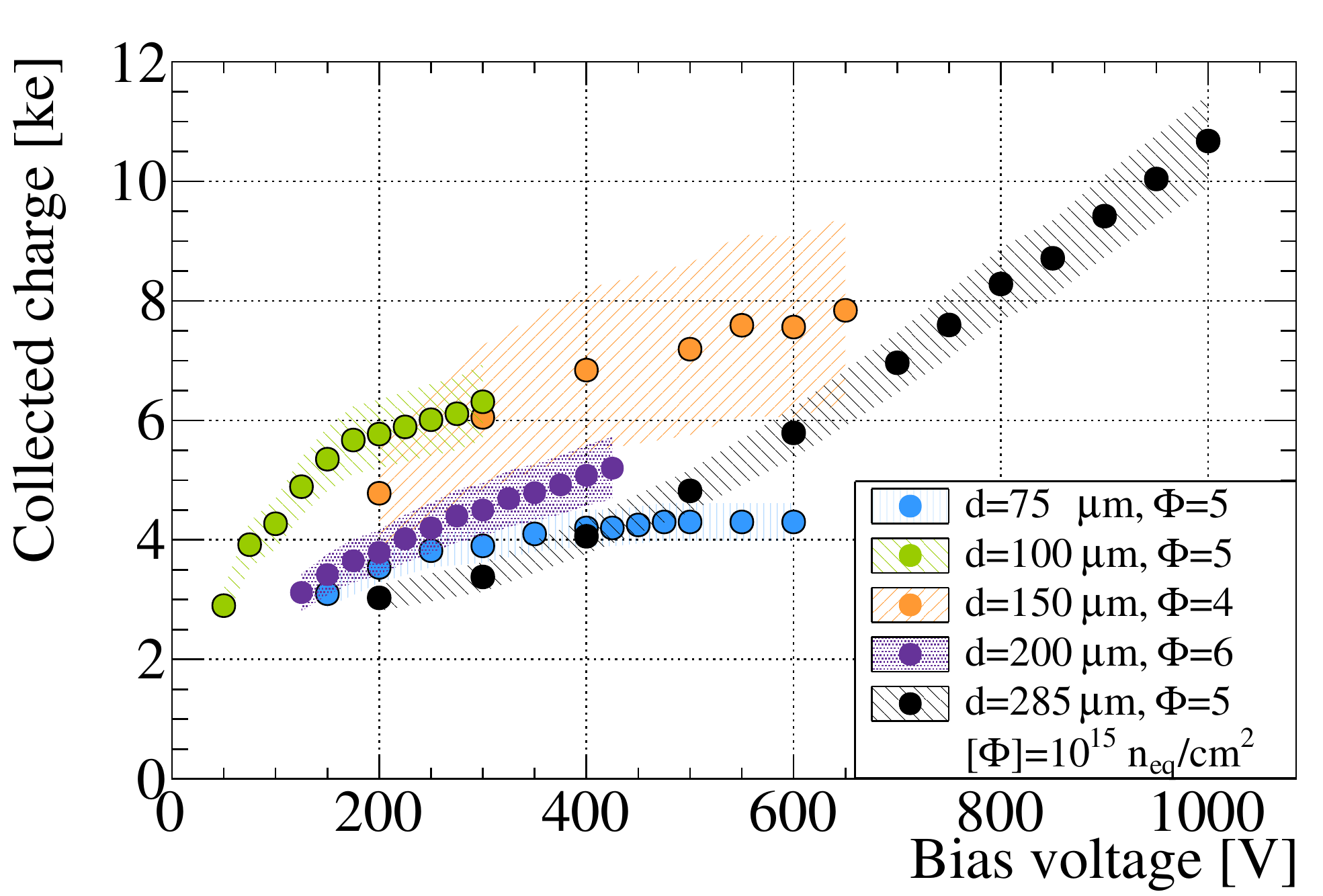}
\caption{}
\label{fig:BT:Thick}
\end{subfigure}
\caption{(a) Inefficiency of n-in-n pixel modules as a function of the bias voltage for various irradiation fluences. Increasing the bias voltage reduces the inefficiency losses. (b) Collected charge as a function of bias voltage for various thicknesses (d) of n-in-p devices all at a fluence of approximately 5~x~10$^{15}$~n$_{eq}$ cm$^{-2}$. In the legend, $\Phi$ represents $10^{15}~n_{eq}~cm^{-2}$.}
\end{figure}
\end{center}

There have been many impressive results for highly irradiated devices. Re-routing of the bias rail, as shown in figure~\ref{fig:BR}, has reduced the inefficiency loss after irradiation for the n-in-p FE-I4 device, KEK46, as shown in figure~\ref{fig:BT:KEK}. At 5~x~10$^{15}$~n$_{eq}$~cm$^{-2}$, the previous design had an efficiency loss of 2--3\% under the bias rail. As illustrated in figure~\ref{fig:BT:ineff}, good efficiency can be reached for irradiated devices at high fluences by increasing the bias voltage. The results in figure~\ref{fig:BT:Thick} show the effect of increasing the bias voltage on the the collected charge for various sensor thicknesses at a fluence of approximately 5~x~10$^{15}$~n$_{eq}$ cm$^{-2}$. At this high fluence with a bias voltage of 200~V, the 100~$\mu$m sample is fully depleted.

\subsection{Large Area Pixel Upgrades}\label{sec:LAPU}

Pixel modules in the outer regions should have a high active area ratio, whilst ideally minimising the contribution to material budget through the overlapping of devices, which has a negative effect on the energy resolution of the calorimeters. Furthermore, migrating to 6'' wafers lowers the cost of production. Consequently prototypes of quad-modules, where one sensor the size of four FE-I4 modules in a 2~by~2 formation is bonded to four FE-I4 chips, have been investigated. Several groups within PPS including Japan, UK, and Germany, have already successfully developed quad-chip modules.

\begin{figure}[tbp] 
\centering
\includegraphics[width=0.7\textwidth]{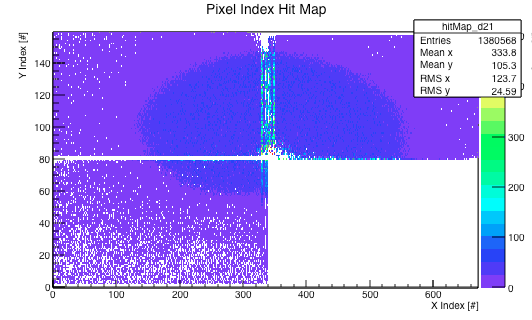}
\caption{Hit efficiency map for n-in-p quad-module, VVTQ5, tested in the SLAC May test beam in 2014. The ganged pixels, which can clearly be seen in the region between two front-end chips, show a lower, but mostly non-zero efficiency.}
\label{fig:LAPU:Quad}
\end{figure}

Figure~\ref{fig:LAPU:Quad} shows efficiency map results from a beam test at SLAC in May 2014 for an n-in-p quad-module known as VVTQ5. The beam was a 12.5~GeV e$^{-}$~beam and the device was tuned to a threshold of 3000 e$^{-}$ with an inverse bias of 100~V applied. The ganged-region between the two top adjacent chips is visible as a lower, but non-zero, efficiency region. The bottom right quarter of the hit efficiency map is empty due to a wire-bonding connection issue with the read-out chip.

\begin{figure}[tbp] 
\centering
\includegraphics[width=0.6\textwidth]{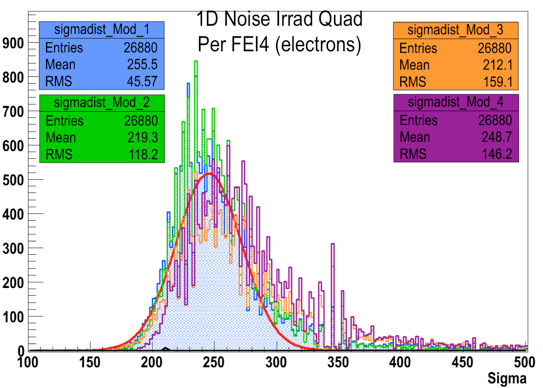}
\caption{Distribution of noise for each of the FE-I4 read-out chips in an irradiated (5~x~10$^{15}$~n$_{eq}$~cm$^{-2}$) quad-module tested in the DESY test beam. The device was characterised to a threshold of 3000 e$^{-}$ and biased with 800~V.}
\label{fig:LAPU:Quad2}
\end{figure}

The results in figure~\ref{fig:LAPU:Quad2} show the noise for each of the four front-end chips on an irradiated quad-module. The module was irradiated to 5~x~10$^{15}$~n$_{eq}$~cm$^{-2}$ and tested at DESY after being characterised to a threshold of 3000 e$^{-}$ and biased with 800~V. The noise, in electrons, is consistent over all chips and the value is as expected.

\subsection{Thin Edgeless Assemblies}\label{sec:TEA}

\begin{figure}[tbp] 
\centering
\includegraphics[width=0.73\textwidth]{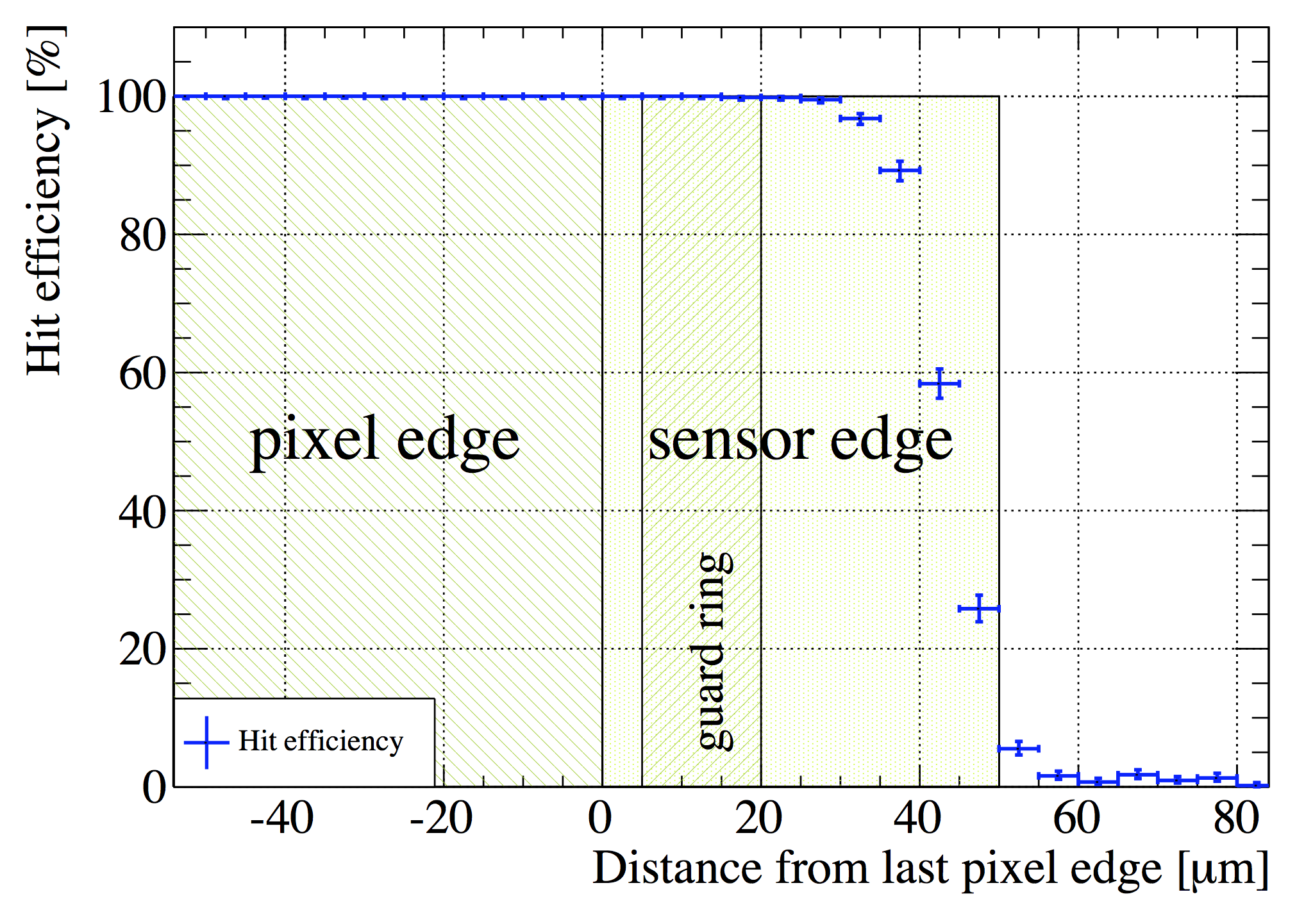}
\caption{One dimensional pixel efficiency map for the edge pixels of the VTT 100~$\mu$m uniradiated FE-I3 device with a 50~$\mu$m active edge. The pixel is $\sim$~99.6\% efficient and the region between the last pixel implant and the module edge (between 0 and +50~$\mu$m) has an efficiency of $\sim$~85$\pm$1~\%.}
\label{fig:TEA}
\end{figure}

As previously discussed, reduction of the sensor edge is required to minimise the dead area of the modules and thus reduce the requirement for shingling. A 100~$\mu$m thick unirradiated FE-I3 device manufactured at VTT, Finland~\cite{ref:VTT}, with 50~$\mu$m active edge has been measured in the November 2013 test beam at DESY~\cite{ref:Terzo}. As can be seen in figure~\ref{fig:TEA}, the efficiency of the edge column for this device (within the region labelled `pixel edge') is very high, at $\sim$~99.6\%. For the area between the last pixel implant and the module edge (within the region labelled `sensor edge') the efficiency is $\sim$~85$\pm$1~\%.

\subsection{Recent SCP Studies}\label{sec:SCP}

The edge of a sensor can be reduced in a process known as Scribing Cleaving Passivation (SCP). The aim is to produce defect-free edges (through cleaving) and then passivate them to keep away the lateral depletion zone. For p-type devices, $Al_2O_3$ can be used for passivation while for n-type devices, silicon oxide and silicon nitride can be used~\cite{ref:SCP}. MOS capacitors with alumina as the dielectric were fabricated to study the effective charge density using C-V curves after irradiation. The $Al_2O_2$ was deposited at two facilities with two different thicknesses (20~nm and 40~nm).

Some of these devices were irradiated with 800~MeV protons at LANSCE up to an equivalent of 0.71~x~10$^{15}$~n$_{eq}$~cm$^{-2}$ (34~Mrad) while others with a gamma source at BNL, receiving a dose of up to 30~Mrad.

\begin{center}
\begin{figure}[tbp] 
\begin{minipage}[c]{.5\textwidth}
\includegraphics[width=\textwidth]{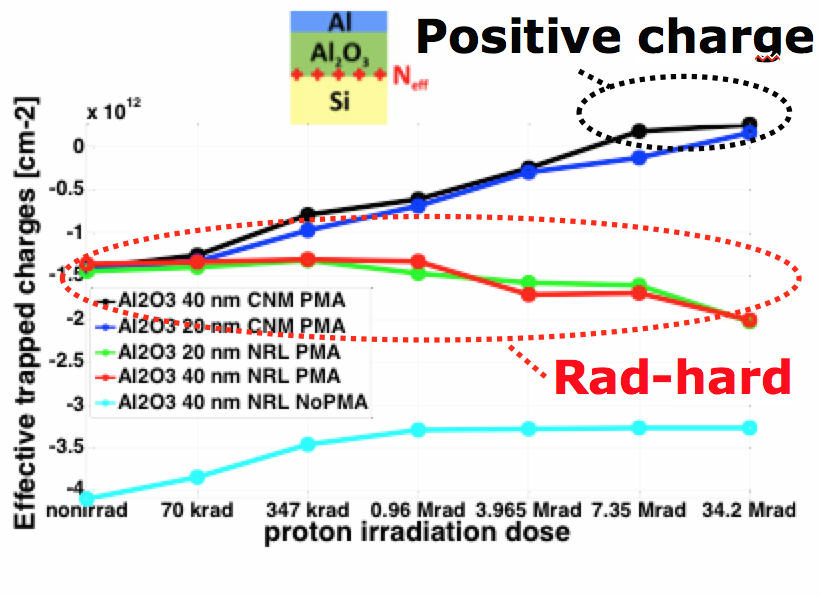}
\end{minipage}
\begin{minipage}[c]{.5\textwidth}
\includegraphics[width=\textwidth]{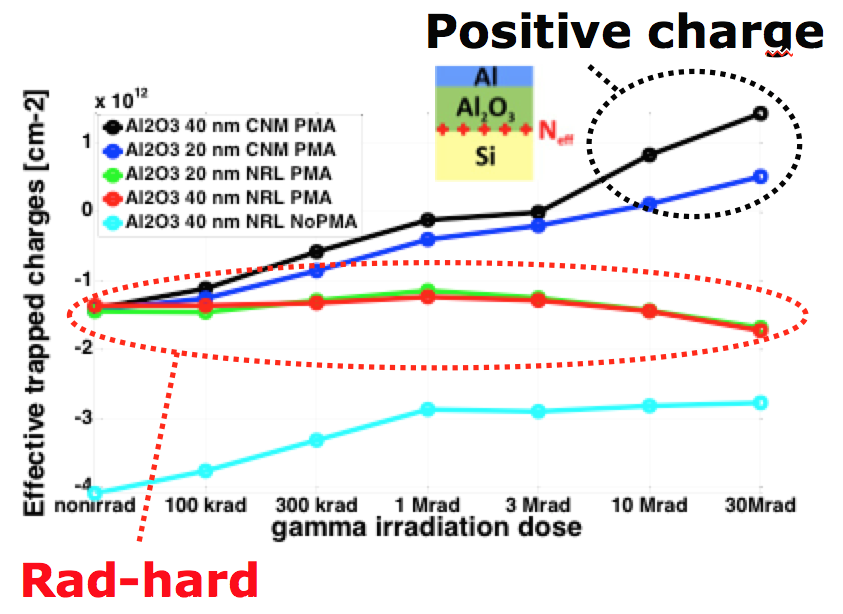}
\end{minipage}
\caption{Effective trapped charge as a function of the proton (left) and gamma (right) irradiation dose of the samples.}
\label{fig:SCP}
\end{figure}
\end{center}

The results of the radiation-induced changes in effective charge density for each process and each irradiation method, are shown in figure~\ref{fig:SCP}. As can clearly be seen, the MOS capacitors processed at the Naval Research Lab~\cite{ref:NRL} (NRL) show a negligible correlation between irradiation dose and the effective trapped charges, while those processed at CNM~\cite{ref:CNM} show trapped charges increasing linearly with dose. Due to these two conflicting results, future fabrications with varied processing are required to further study the processing differences and their effect on the radiation performance.

\subsection{Doping Profile Measurements}\label{sec:DPM}

\paragraph{Motivations for producing simulations}

Results obtained from test structures can be used to develop reliable simulations of devices which, in turn, drive the development of new sensor layouts. This process is usually significantly quicker and less expensive than building multiple prototypes. Therefore, the validation of simulation models is vital to trust predictions.

\paragraph{Secondary Ion Mass Spectrometry}

The process of Secondary Ion Mass Spectrometry (SIMS) is an analytical technique that is used to characterise the various impurities in the surface and near surface region of a material. In this case a depth of up to 30~$\mu$m from the surface is analysed. The SIMS process uses the sputtering of a primary energetic ion beam, with an energy between 0.5--20~keV, on the surface of a sample and consequently analyses mass spectrum of the ionised secondary particles that are produced. The process is destructive, leaving a crater in the sample, and hence, samples are specifically produced for this analysis technique. Through this method, the total dopant density of a sample can be determined~\cite{ref:SIMS}.

\begin{center}
\begin{figure}[tbp] 
\begin{minipage}[c]{.485\textwidth}
\includegraphics[width=\textwidth]{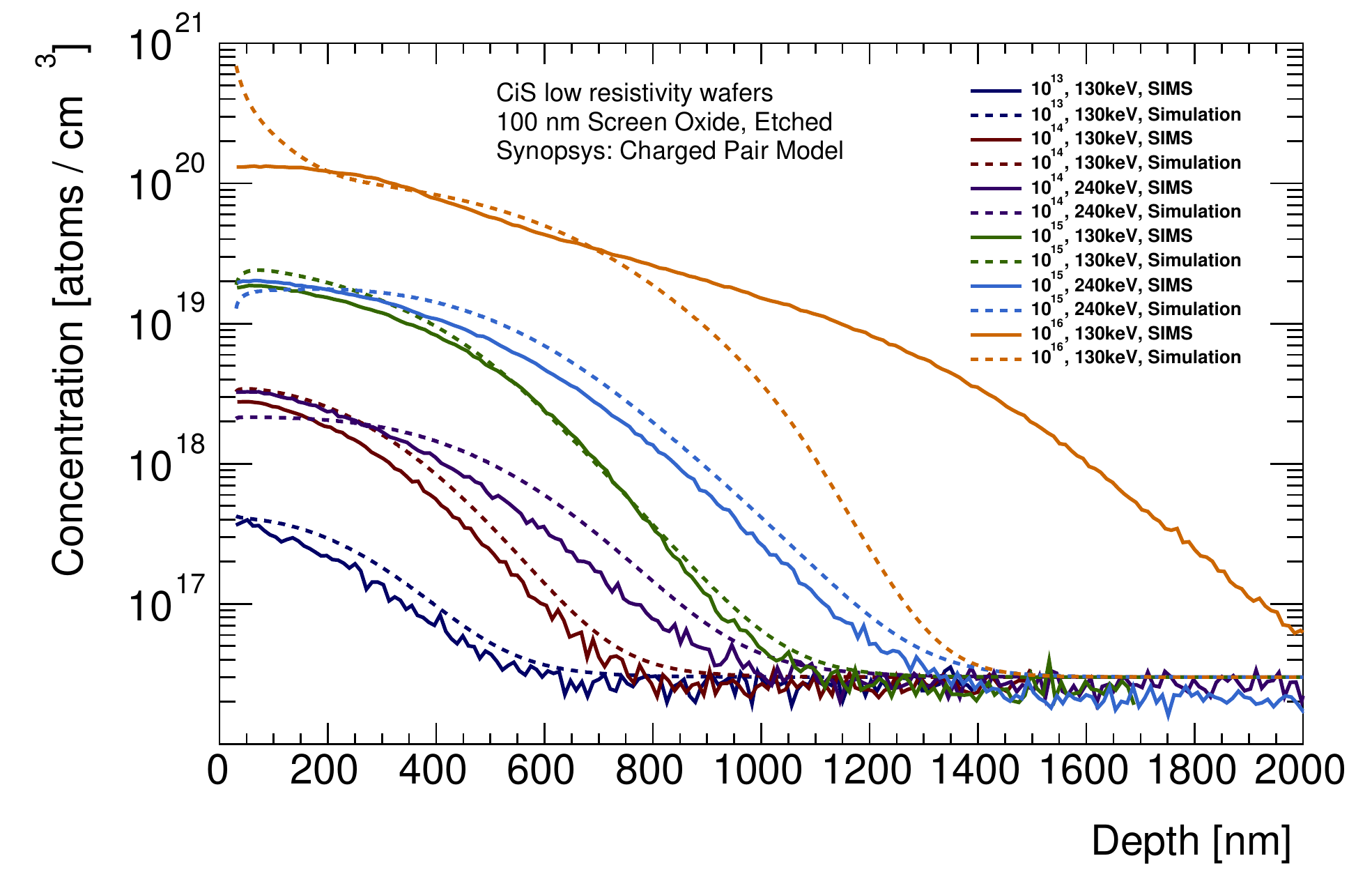}
\end{minipage}
\begin{minipage}[c]{.03\textwidth}
\end{minipage}
\begin{minipage}[c]{.485\textwidth}
\includegraphics[width=\textwidth]{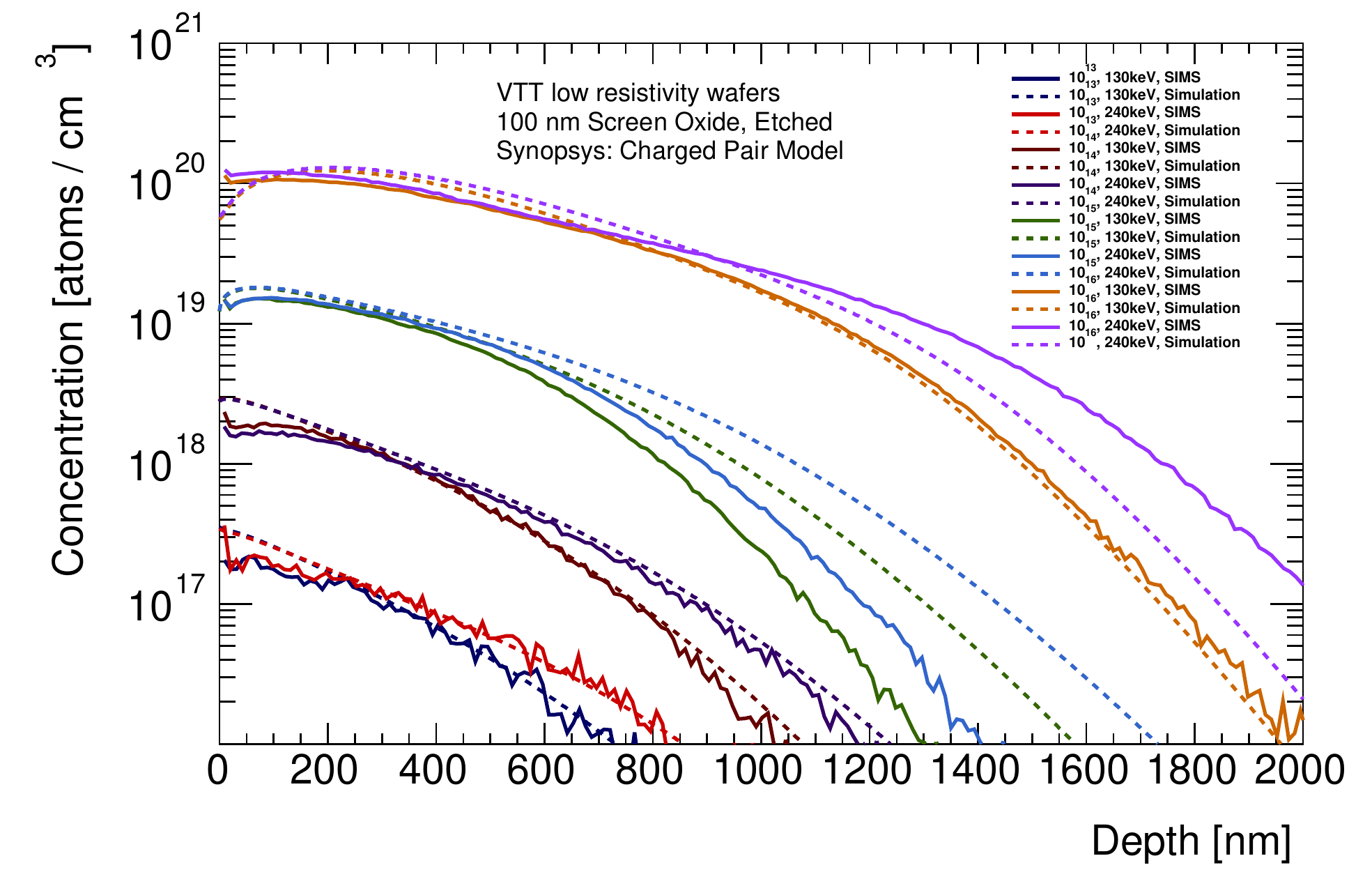}
\end{minipage}
\caption{SIMS measurements of samples at various doping profiles compared to TCAD Synopsys simulations using the Charged Pair diffusion model. The samples were manufactured by CiS (left, see table~1) and VTT (right, see table 2).}
\label{fig:SIMS}
\end{figure}
\end{center}

The comparison of Technology Computer Aided Design (TCAD) simulation and SIMS data for devices manufactured at CiS, Germany~\cite{ref:CIS} and at VTT are shown in figure~\ref{fig:SIMS}. The diffusion model used for the simulation is the Synopsys~\cite{ref:Synop} Charged Pair simulation model. The samples used for the study are shown in table~\ref{table:SIMS-CiS} for the n-in-n samples manufactured at CiS, and table~\ref{table:SIMS-VTT} for the n-in-p samples manufactured at VTT. 

There is generally good agreement between the simulation and SIMS data for lower doses, however, there is a large discrepancy with the highest dose (implantation dose of 10$^{16}$) for the sample from the CiS production; the reason for this is currently not completely understood. A further comparison with various diffusion models in Synopsys and Silvaco~\cite{ref:Silvo} has been performed, but currently no single model accounts for all of the features of the SIMS data. The samples manufactured at VTT were subjected to a different annealing process that lead to the oxidation of the surface region (later removed), which was accounted for in the simulation. Given the removal of surface impurities, the closer agreement in the surface region for these samples is expected.

\begin{table}[tbp]
\begin{center}
\caption{Samples prepared with various implantation doses and energies and subsequently analysed with the Secondary Ion Mass Spectrometry process to study the total doping profile in the near surface region. The samples were manufactured at CiS.} 

\begin{tabular}{ l | c | c | c | c | c | c | c | c }
\hline
\multicolumn{9}{c}{n-in-n, CiS production, \textless100\textgreater~orientation, thickness 380~$\mu$m} \\ \hline \hline
Oxide thickness & \multicolumn{8}{c}{100 nm} \\
P implantation doses & \multicolumn{2}{c |}{10$^{13}$~cm$^{-2}$} & \multicolumn{2}{c |}{10$^{14}$~cm$^{-2}$} & \multicolumn{2}{c|}{10$^{15}$~cm$^{-2}$} & \multicolumn{2}{c}{10$^{16}$~cm$^{-2}$} \\
Implantation energy (keV) & 130 & 240 & 130 & 240 & 130 & 240 &  \multicolumn{2}{c}{130}  \\
Annealing & \multicolumn{8}{c}{4~hours, 975~$^{\circ}$C} \\ \hline
\end{tabular}

\label{table:SIMS-CiS}
\end{center}
\end{table}

\begin{table}[tbp]
\begin{center}
\caption{Samples prepared with various implantation doses and energies and subsequently analysed with the Secondary Ion Mass Spectrometry process to study the total doping profile in the near surface region. The samples were manufactured at VTT.} 

\begin{tabular}{ l | c | c | c | c | c | c | c | c }
\hline
\multicolumn{9}{c}{n-in-p, ADVACAM production, \textless100\textgreater~orientation, thickness $\leq$ 675~$\mu$m} \\ \hline \hline
Oxide thickness & \multicolumn{8}{c}{100 nm} \\
P implantation doses & \multicolumn{2}{c |}{10$^{13}$~cm$^{-2}$} & \multicolumn{2}{c |}{10$^{14}$~cm$^{-2}$} & \multicolumn{2}{c |}{10$^{15}$~cm$^{-2}$} & \multicolumn{2}{c}{10$^{16}$~cm$^{-2}$} \\
Implantation energy (keV) & 130 & 240 & 130 & 240 & 130 & 240 & 130 & 240  \\
Annealing & \multicolumn{8}{c}{3~hours, 1000~$^{\circ}$C} \\ \hline
\end{tabular}
\label{table:SIMS-VTT}
\end{center}
\end{table}

\section{Summary and Conclusion}\label{sec:summary}

This paper has presented the recent achievements of the Planar Pixel Sensor group, as part of the ATLAS ITk Pixel Collaboration, working towards improved pixel devices for the Phase II upgrade of the ATLAS inner detector.

Planar sensors can be operated up to 2 x 10$^{16}$~n$_{eq}$~cm$^{-2}$ with good efficiency. An optimisation of the pixel cell design, especially in the bias grid structures is promising for a further improvement of the performance after irradiation. Several methods for edge slimming have proven to work and now the focus is on optimising the cost of these methods. Indeed, substantial cost reduction compared to current pixel detector has been achieved, also by adopting n-in-p design on 6'' wafers; 8'' wafer production is currently being assessed. The first quad-modules, designed for the outer-layers, have been prototyped by several member institutes and have successfully been operated after irradiation in a beam test. Comparison of SIMs measurements and TCAD simulation have shown good agreement for lower doses and further study of diffusion models for higher dose measurements is required.

\acknowledgments

This work has been partially performed in the framework of the CERN RD50 Collaboration. The authors thank V. Cindro for the irradiation at JSI, A.~Dierlamm for the irradiation at KIT and S. Seidel (University of New Mexico) for the irradiations at LANSCE. The irradiations at KIT were supported by the Initiative and Networking Fund of the Helmholtz Association, Contract HA-101 (Physics at the Terascale). The irradiation at JSI and the beam-tests have received funding from the European Commission under the FP7 Research Infrastructures project AIDA, Grant agreement no. 262025.

\end{document}